\documentclass[runningheads]{svmult}
\usepackage{makeidx}   
\usepackage{graphicx}  
\usepackage{epsfig}
\usepackage{amsmath}
\usepackage{amssymb}
\usepackage{subeqnar}  
\usepackage{multicol}  
\usepackage{cropmark} 
\usepackage{physprbb}  
\makeindex             
\begin{document}
\title*{Soft Limits of Multiparticle Observables and
Parton Hadron Duality
\footnote{
Presented at Ringberg Workshop ``New trends in HERA physics 1999'',
Tegernsee, Germany, May 30 - June 4, 1999}
}
%
%
\toctitle{Soft Limits of Multiparticle Observables and
Parton Hadron Duality}
%
%
\titlerunning{Soft Limits of Multiparticle Observables}
%
\author{Wolfgang Ochs\inst{}}
%
%
%
\institute{Max-Planck-Institut f\"ur Physik (Werner Heisenberg-Institut),
F\"ohringer Ring 6, D-80805 M\"unchen, Germany}

\maketitle              

\begin{abstract}
We discuss observables 
in multiparticle production for three kinds of limits of decreasing kinematical
scales:
1. the transition jet $\to$ hadron (limit $y_{cut}\to 0$
of the resolution parameter $y_{cut}$); 2. single particle inclusive 
distributions normalized at threshold $\sqrt s\to 0$ and 3. particle 
densities in the limit of low momentum $p,p_T\to 0$. The observables 
show a smooth behaviour in these limits and 
follow the perturbative QCD predictions, originally designed
for large scales, whereby a simple prescription is supplemented 
to take into account
mass effects. A corresponding physical picture is described.
\end{abstract}

\section{Introduction}
A successful description of multiparticle production
based on perturbative QCD has been established for "hard"
processes which are initiated by an interaction of 
elementary quanta (quarks, leptons, gauge bosons, ...) at large momentum
transfers $Q^2\gg\Lambda^2$, whereby the 
characteristic scale in QCD is $\Lambda\sim$ few 100 MeV.
In this kinematic regime the  running coupling constant
$\alpha_s(Q^2) $ is small and the lowest order terms 
of the perturbative expansion provide the desired accuracy.  
The coloured quarks and gluons which emerge from the primary hard 
process cannot escape towards large distances because of the confinement of
the colour fields. Rather, they ``fragment'' into particle jets 
which may consist of many stable and unstable hadrons.   

Here we are interested in the emergence of the hadronic final
states and jet structure. The partons participating in the
hard process generate  parton cascades through gluon Bremsstrahlung and quark
antiquark pair production processes which
can be treated again perturbatively, at least approximately.
 The  singular behaviour of the gluon Bremsstrahlung 
in the angle $\Theta$ and momentum~$k$
\begin{equation}
\frac{dn}{dk d\Theta}\propto 
\alpha_s(k_T/\Lambda)\frac{1}{k\Theta}, 
\quad k_T>Q_0.
\label{brems}
\end{equation}
(in  lowest order and for small angles)
leads to the collimation of the partons and the jet structure.
The transverse momentum $k_T$
is taken as characteristic scale for the
coupling  $\alpha_s\sim 1/\ln(k_T/\Lambda)$,
so it will rise with decreasing scale during jet evolution
 and one expects the
perturbation theory to loose its valididity below a limiting scale $Q_0$. 

The  transition to the hadronic final state, finally, proceeds
at small momentum transfers $k_T\sim Q_0$ by
non-perturbative processes. 
There have been different approaches to obtain predictions on 
 the hadronic final states:
\enlargethispage{1.0cm}
\pagebreak

\noindent {\it 1. ``Microscopic'' Monte Carlo models}\\
In a first step a parton final state 
is generated perturbatively corresponding to a cut-off scale like the 
above $Q_0$. Then, according to a non-perturbative model
intermediate hadronic systems (clusters, strings, \ldots) are formed 
which decay, partly through intermediate resonances, into the final hadrons
of any flavour composition.  
Depending on the considered complexity a larger number
of adjustable parameters are allowed for in addition to the QCD 
scale and cut-off parameters.
 Because of the complexity of these models only Monte Carlo methods
are available for their analysis. They are able to reproduce many very
detailed properties of the final state successfully.

\noindent {\it 2. Parton  Hadron Duality approaches}\\
One compares the perturbative QCD result for particular observables directly
with the corresponding result for hadrons. The idea is that the effects of
hadronization are averaged out for sufficiently inclusive observables.
In this case analytical results are aimed for which are closer to a direct
physical interpretation than the MC results (for reviews, see
\cite{dkmt2,ko}).
This general idea comes in
various realizations, we emphasize three kinds of observables:

\noindent {\it Jet cross sections:} 
Jets are defined with respect to a 
certain resolution criterion (parameter $y_{cut}$), then 
the cross sections for hadron and parton jets are compared directly
at the same resolution. This
phenomenological ansatz has turned out to be extremely successful in 
the physics of energetic jets.
A priori, it is nontrivial that  
an energetic hadron jet with dozens of hadrons 
 should be compared directly to a parton jet
with only very few (1-3) partons.\\
\noindent {\it Infrared and collinear safe observables:} 
The value of such an observable is not changed if 
a soft particle with $k\to 0$ or a collinear particle ($\Theta\to 0$) is
added to the final state. It is then expected that the observables are less
sensitive to the kinematic region $k_T\sim Q_0$ in (\ref{brems}). 
Especially, event shape observables like "Thrust" 
or energy flow patterns
belong into this 
category. Perturbative calculations with all order resummations
have been generally successful. In recent years
perturbative calculations to $O(\alpha_s^2)$ in combination with  
 power corrections $\sim 1/Q^q$ have found considerable interest.\\
\noindent
{\it Infrared sensitive observables:} 
Global particle multiplicities as well as inclusive
particle distributions and
correlations belong into this category; these observables are divergent 
for $Q_0\to 0$ and therefore are
particularly sensitive to the transition region from partons to hadrons.
$Q_0$ plays the role of a nonperturbative hadronization parameter.

In this report we will be concerned with the last class of observables
to learn about the soft phenomena and ultimately about the colour
confinement mechanisms.
Specific questions concerning the role of perturbative QCD are
\begin{itemize}
\item What is the limiting value of $Q_0$ for which perturbative QCD can be
applied successfully. Especially, can $Q_0$ be of the order of $\Lambda\sim$
few 100 MeV?
\item Is there any evidence for
the strong rise of the coupling constant $\alpha_s$ towards small scales
below 1 GeV?
\item Is there evidence for characteristic QCD coherence effects at small scales
which are expected for soft gluons, evidence for the colour factors
$C_A,C_F$?
\end{itemize}

\section{Theoretical approach}
\subsection{Partons}
The evolution of a parton jet is described in terms of a multiparticle
generating functional $Z_A(P,\Theta;\{u(k)\})$ with momentum test functions
$u(k)$ for a primary parton $A$
($A=q,g$) of momentum $P$ and jet opening angle $\Theta$. This functional
fulfils a differential-integral equation
 \cite{dkmt2}
\begin{equation}
\begin{split}
\frac{d}{d \: \ln  \Theta} \: Z_A (P,& \Theta)  =
\frac{1}{2}
\; \sum_{B,C} \; \int_0^1 \; dz  \\
 \; &\times \ \frac{\alpha_s (k_T)}{2 \pi} \: \Phi_A^{BC} (z)
\left [Z_B (zP, \Theta) \: Z_C ((1 - z)P, \Theta) \: - \: Z_A
(P,\Theta) \right ]
  \label{Zevol}
\end{split}
\end{equation}
and has to be solved with constraints $k_T>Q_0$ and with initial condition  
\begin{equation}
Z_A (P, \Theta; \{ u \})|_{P \Theta  =  Q_0} \; = \; u_A
(k =  P).
\label{init}
\end{equation}
which means that at threshold $P \Theta =  Q_0$ there is only one particle
in the jet.
From the functional $Z_A$ one can obtain the inclusive n-parton momentum 
distributions by functional differentiation after the functions $u(k_i)$,
$i=1\ldots n$, at
u=1 and then one finds the corresponding evolution equations as in (\ref{Zevol}).
This ``Master Equation'' includes the following features:
the splitting functions $\Phi_A^{BC} (z)$ 
of partons $A\to BC$;
evolution in angle $\Theta$ yielding
a sequential angular ordering 
 which limits the phase space of soft emission as a consequence
of colour coherence; the running coupling $\alpha_s(k_T)$.
For large 
momentum fractions $z$ the equation approaches the 
usual DGLAP evolution equations.
 
The solution of the evolution equations can be found by iteration and then
generates an all order perturbation series; it is
complete in leading order (``Double Logarithmic Approximation -- DLA)
 and in the
next to leading order (``Modified Leading Log Approximation'' -- MLLA),
 i.e. in the terms
$\alpha_s^n \log ^{2n}(y)$ and $\alpha_s^n \log ^{2n-1}(y)$. The
logarithmic terms of lower order
are not complete, but it makes sense to include them as
well as they are important for taking into acount energy conservation and
the correct behaviour near threshold (\ref{init}).
The complete partonic
final state of a reaction may be constructed by matching with 
an exact matrix element result for the primary hard process.

\subsection{Hadrons}
We investigate here the possibility that the parton cascade
resembles the hadronic final state for
sufficiently inclusive quantities. One motivation is ``preconfinement''
\cite{preconf},
the preparation of colour neutral clusters of limited mass within the
perturbative cascade. 
If the cascade is evolved towards a low scale $Q_0\sim\Lambda$,
a successful description 
of inclusive single particle distributions has been obtained
(``Local Parton Hadron Duality''-LPHD \cite{adkt1}). More generally,
one could test relations between parton and hadron observables of the
type 
\begin{equation}
O(x_1,x_2,...)|_{hadrons}\; = \;K\: O(x_1,x_2,...;Q_0,\Lambda)|_{partons}
\label{lphdeq}
\end{equation}
where the nonperturbative cut-off $Q_0$ and an arbitrary factor $K$ are
to be determined by experiment (for review, see \cite{ko}). 
In comparing differential parton and hadron distributions there can be a
mismatch near the soft limit because of mass effects, especially, 
the (massless) partons 
are restricted by $k_T>Q_0$ in (\ref{Zevol})
but hadrons are not. This mismatch can be avoided by an appropriate choice of
energy and momentum variables. In a simple model  \cite{lo,klo} one compares
partons and
hadrons  at the same energy (or transverse mass) using an
effective mass $Q_0$ for the hadrons, i.e. 
\begin{equation}
E_{T,parton}=k_{T,parton}\qquad \Leftrightarrow \qquad
E_{T,hadron}=\sqrt{k_{T,hadron}^2+Q_0^2},  \label{phcorr} 
\end{equation}
then, the corresponding lower limits are $k_{T,parton}\to Q_0$ and 
 $k_{T,hadron}\to 0$.

\section{From  Jets to Hadrons, the limit $y_{cut}\to 0$}

We turn now to the discussion of several observables
and their behaviour in the limit of a small scale. 
First, we consider the transition from jets to hadrons by decreasing the
resolution scale of jets. Jet physics is a standard testing ground for
perturbative QCD, the transition to hadrons 
therefore corresponds to the transition from the known to the unknown
territory.

The jets are defined in the multiparticle final state by a
cluster-algorithm. Popular is the ``Durham algorithm'' \cite{durham}
which allows the all order summation in the perturbative analysis. For a
given resolution parameter $y_{cut}=(Q_{cut}/Q)^2$ in a final state with
total energy $Q$ particles are successively combined into clusters until all
relative transverse momenta are above the resolution parameter
 $y_{ij}=k_T^2/Q^2>y_{cut}$.\footnote{more precisely, the distance is defined
by  $y_{ij}= 2(1-\cos\Theta_{ij})\ {\rm
min}(E_i^2E_j^2)/Q^2>y_{cut}$.
}
We study now the mean jet multiplicity $N_{jet}$ in the event as function of
$y_{cut}$.  In $e^+e^-$-annihilation for $y_{cut}\to 1$ all particles are
combined into two jets and therefore $N_{jet}=2$, on the other hand, for
$y_{cut}\to 0$ all hadrons are resolved and $N_{jet}\to N_{had}$.
%
\begin{figure*}[t]
\begin{center}
\mbox{\epsfig{file=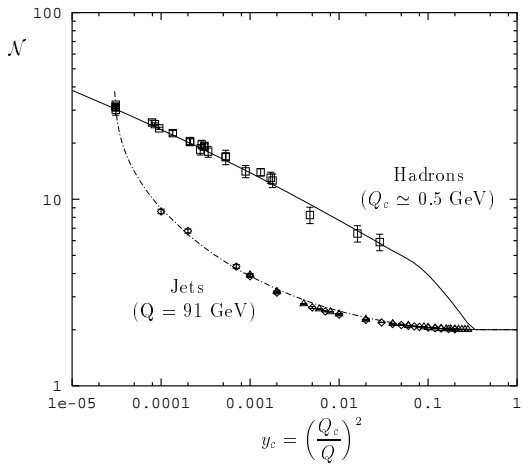,width=7.0cm,bbllx=3.2cm,bblly=9.2cm,bburx=18.cm,bbury=22.8cm}}
\end{center}
\vspace{-0.7cm}
\caption[]{
Data on the average jet multiplicity $\protect {\mathcal N}$ at $Q$ = 91 GeV 
for different resolution parameters $y_c$ (lower set) and
the average hadron multiplicity (assuming $\protect {\mathcal N} = \frac{3}{2}
\protect {\mathcal N}_{ch}$)
at different $cms$ energies between $Q=3$ and $Q=91$ GeV using
$Q_c=Q_0$ = 0.508 GeV in the parameter $y_c$ (upper set).
The  curves follow from the evolution equation (\protect\ref{Zevol})
with $\Lambda$ = 0.5 GeV; the upper curve for hadrons is based on 
the duality picture (\ref{lphdeq}) with $K=1$ and parameter $Q_0$
(Fig. from \cite{lo2}) }
\label{fig1}
\end{figure*}
\begin{figure*}[ht]
\begin{center}
\noindent
\begin{minipage}{5.8cm}
\mbox{\epsfig{bbllx=0bp,bblly=45bp,bburx=285bp,bbury=280bp,%
file=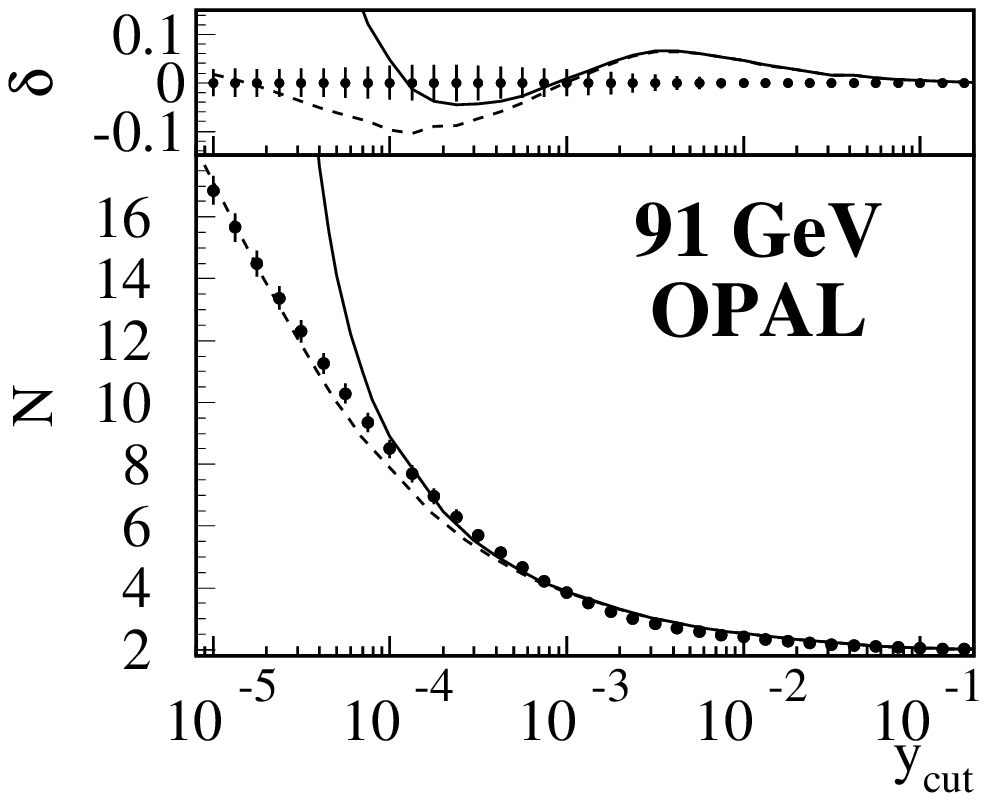,
width=5.8cm}}
\end{minipage}
\hfill 
\begin{minipage}{5.8cm} 
\mbox{\epsfig{bbllx=0bp,bblly=45bp,bburx=285bp,bbury=280bp,%
file=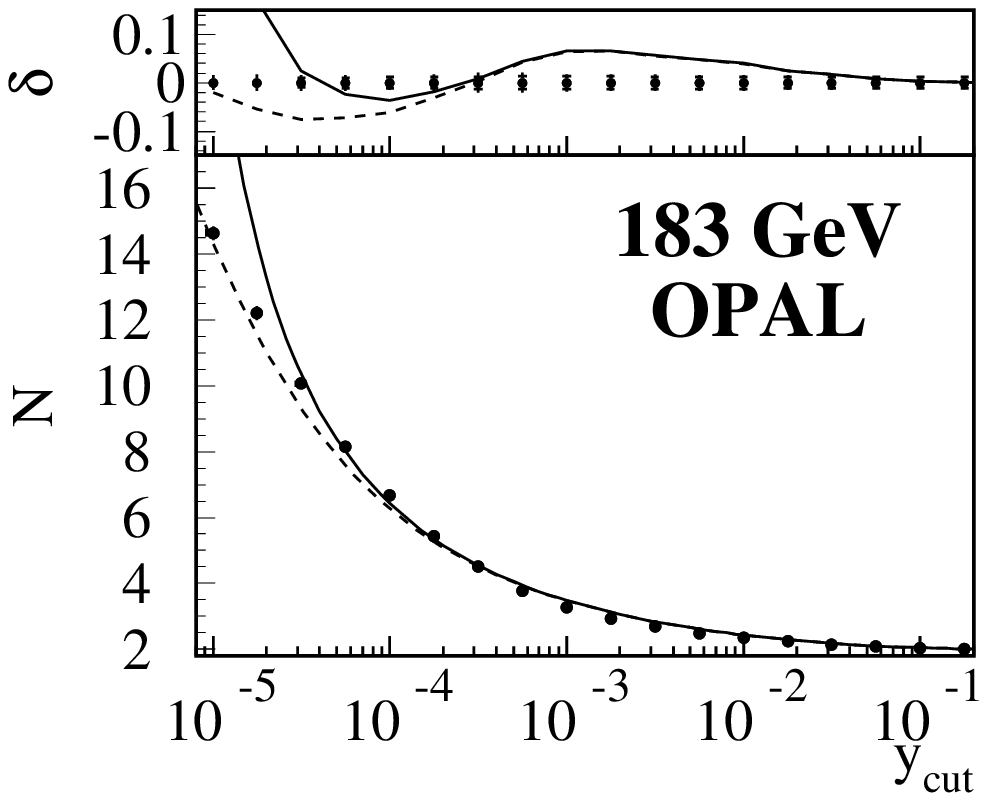,
width=5.8cm}}
\end{minipage}
\end{center} 
\vspace{-0.3cm}
\caption{Jet multiplicities 
extending towards lower $y_{cut}$ parameters; full lines as in
Fig.~1 for jets, dashed lines the same predictions but shifted $y_{cut}\to
y_{cut}-Q_0^2/Q^2$ according to the different kinematical boundaries  
as in (\ref{phcorr}), with parameters as in Fig. 1 
(preliminary data from OPAL \cite{Pfeifenschneider})}
\label{fig2a}
\end{figure*}

Results on jet multiplicities are shown in Fig. 1. 
 The jet multiplicity rises only slowly 
with decreasing
$y_{cut}$. For $y_{cut} \gtrsim 0.01$ the data are well described by
the complete matrix element calculations  to $O(\alpha_s^2)$ 
(first results of this kind in \cite{kl}) 
and allow the precise determination of the
coupling or, equivalently, of the QCD scale parameter
$\Lambda_{\overline{MS}}$
\cite{L3jmul,opaljmul}. In the region above $y_{cut}>10^{-3}$ 
the resummation of the higher orders in $\alpha_s$ becomes important
\cite{cdotw} and the MLLA calculation describes the data well.
The lower curve shown in Fig. 1 is obtained \cite{lo2} from a full 
(numerical) solution of the evolution equations corresponding to
(\ref{Zevol}), matched with the $O(\alpha_s)$ matrix element,
 and describes the data  obtained at LEP-1
\cite{L3jmul,opaljmul} down to $10^{-4}$.

The theoretical curve diverges for small cut-off  $Q_{cut}\to \Lambda$
as in this case the coupling $\alpha_s(k_T)$ diverges. 
In the duality picture discussed above the parton final state corresponds to
a hadron final state at the resolution $k_T\sim Q_0$ according to
(\ref{lphdeq}) and this limit is reached for $Q_{cut}\to Q_0$.
The calculation meets the hadron multiplicity data  
for the cut-off parameter $Q_0\simeq 0.5$ GeV.  If this
calculation is done for lower $cms$ energies, 
agreement with all hadron multiplicity 
data down to $Q=3$ GeV is obtained with  the same parameter $Q_0$
as seen in Fig. 1 by the upper set of data and the theoretical curve.
Moreover, the normalization constant in (\ref{lphdeq}) can be chosen as
$K=1$ whereas in previous approximate calculations $K\approx 2$
(see, e.g. \cite{lo}). This result 
implies that the hadrons, in the duality picture, correspond
to very narrow jets with resolution $Q_0\simeq 0.5$ GeV.

In this unified description of hadron and jet multiplicities the running of
the coupling plays a crucial role. Namely, for constant $\alpha_s$ both
curves for hadrons and jets in Fig. 1 would coincide, as only one scale
$Q_{cut}/Q$ were available. With running $\alpha_s(k_T/\Lambda)$ the absolute
scale of $Q_{cut}$ matters: $\alpha_s$ varies most strongly for 
$Q_{cut}\to\Lambda$ for jets at small $y_{cut}$ in the transition to hadrons
and for hadrons near the threshold of the process at large  $y_{cut}$
where $\alpha_s>1$. 
It appears that the final stage of hadronization in the jet evolution can be
well represented by the parton cascade with the strongly rising 
 coupling.    

Preliminary results on jet multiplicities at very small $y_{cut}$
have been obtained recently by
OPAL \cite{Pfeifenschneider} and examples 
are shown in Fig. 2. Whereas in the theoretical calculation all hadrons 
 (partons in the duality picture) are resolved 
for $Q_{cut}\to Q_0$, 
for the experimental quantities this limit occurs for $Q_{cut}\to 0$.

This is an example of the kinematical mismatch between
experimental and theoretical quantities discussed above 
and can be taken into account \cite{lo2} by a shift in $y_{cut}$ 
according to (\ref{phcorr}). The shifted (dashed) curves in Fig. 2
describe the data rather well (also at intermediate $cms$ energies)
 whereby the $Q_0$ parameter has been taken from
the fit to the hadron multiplicity before; the predictions fall a bit 
below the data at lower energies like 35 GeV. The 
nonperturbative $Q_0$ correction becomes negligable for $Q_c\gtrsim 1.5$
GeV.

 We conclude that in case of this simple global observable the perturbative
QCD calculation provides a good description of hard and soft phenomena
in terms of one non-perturbative
parameter $Q_0\sim \Lambda$ (from fit \cite{lo2}  
$Q_0\approx 1.015 \Lambda$).
Multiplicity moments are described very well in this approach also
\cite{lupia}. 

\section{Shape of Energy Spectrum, the Limit $\sqrt{s}\to 0$}

A standard procedure in perturbative QCD is the derivation of the $Q^2$
evolution of the inclusive distributions -- either of the structure
functions in DIS ($Q^2<0$) or of the hadron momentum 
distributions (``fragmentation
functions'', $Q^2\equiv s>0$). One starts from an input function at an initial scale 
$Q_1^2$ and predicts the change of shape with $Q^2$.   

In the LPHD picture  one derives the parton distribution from the evolution
equation (\ref{Zevol}) with initial condition (\ref{init}) at
threshold, here the spectrum is simply 
\begin{equation}
D(x,Q_0)=\delta(x-1). \label{Dthresh}
\end{equation}
If we start from this initial condition
the further QCD evolution
predicts the absolute shape of the particle 
energy distribution at any higher $cms$ energy $\sqrt{s}$. Within certain
high energy approximations one can let $Q_0\to \Lambda$ and obtains an
explicit analytical expression for the spectrum
in the variable $\xi=\ln(1/x)$, the so-called ``limiting
spectrum'' \cite{adkt1} which has been found to agree well with the data 
in the sense of (\ref{lphdeq}) --
disregarding the very soft region $p\lesssim Q_0$ (see, e.g. the 
review \cite{ko}). 
In the more general case $Q_0\neq \Lambda$ the cumulant
moments $\kappa_q$ of the $\xi$ distribution
have been
calculated as well \cite{FW,DKTInt};
they are defined by $\kappa_1 = <\xi> = \bar
\xi$, $\kappa_2 \equiv \sigma^2 = <(\xi - \bar \xi)^2>$, $\kappa_3 = <(\xi -
\bar \xi)^3>$, $\kappa_4 = <(\xi - \bar \xi)^4> - 3 \sigma^4$, \dots;
also one introduces the reduced cumulants $k_q \equiv \kappa_q/ \sigma^q$,
in
particular the skewness $s = k_3$ and the kurtosis $k = k_4$.

In the comparison with data some attention has to be paid again to the soft
region. The experimental data are usually presented in terms of the momentum
fraction $x_p=2p/\sqrt{s}$, then $\xi_p\to\infty$ for $p\to 0$. On the other
hand, the theoretical distribution, because of $p>p_T>Q_0$, is limited
to the interval $0<\xi<Y$, $Y=\ln(\sqrt{s}/2Q_0)$. Therefore, in this region
near and beyond the boundary the two  distributions cannot agree. 
A consistent description can be obtained if theoretical and experimental
distributions are compared at the same energy 
as in (\ref{phcorr}),
then both $\xi$ spectra have the
same upper limit $Y$. With a corresponding ``transformation'' of
$E\frac{d^3n}{d^3p}$ the 
spectra are well described by the appropriate
theoretical formula near the boundary \cite{lo}.

\begin{figure}
          \begin{center}
\vspace{-3.3cm}
      \mbox{  \hspace{-1.5cm}
\mbox{\epsfig{file=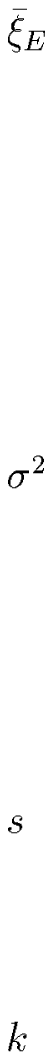,bbllx=1.0cm,bblly=7.cm,bburx=5.2cm,bbury=26.cm,%
height=12cm}} \hspace{0.2cm}
\mbox{\epsfig{file=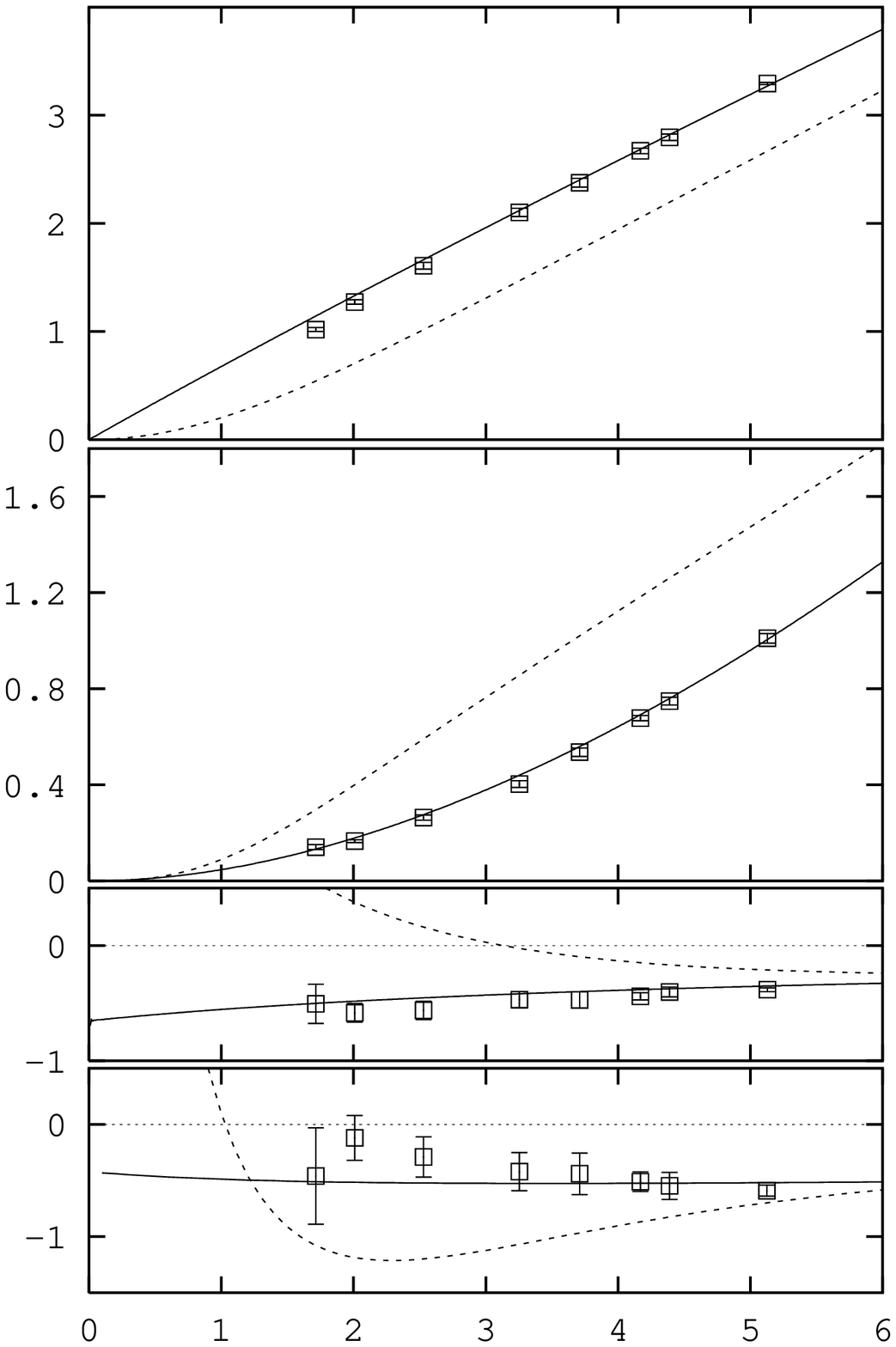,bbllx=5cm,bblly=7cm,bburx=18.cm,bbury=27.5cm,%
height=8.4cm}}
}          \end{center}
\vspace{-0.2in}
\mbox{\hspace{5.2cm} $Y = \log (\sqrt{s}/2 Q_0)$}
\vspace{0.3cm}
\caption{
The first four
cumulant moments of charged particles' energy spectra 
i.e., the average   
value $\bar \xi_E$, the dispersion $\sigma^2$, the skewness $s$ and the
kurtosis $k$,       
are shown as a function of $cms$ energy $\sqrt{s}$ 
for $Q_0$ = 270 MeV and  $n_f$ = 3, in comparison with
MLLA predictions of the ``limiting spectrum'' (i.e. $Q_0 = \Lambda$)
for running $\alpha_s$ (full line) 
 and for fixed $\alpha_s$ (dashed line) (from \cite{lo})
}
\label{fig:moments}
\end{figure}

The cumulant moments of the energy spectrum of
hadrons determined in this way have been compared \cite{lo} 
with the theoretical calculation based on the MLLA evolution
equation  \cite{DKTInt}. As seen in Fig. \ref{fig:moments}
the data agree well
with the limiting spectrum result  ($Q_0=\Lambda$), 
both in their energy
dependence and their absolute normalization at threshold (the moments 
vanish because of (\ref{Dthresh})).
This suggests that perturbative calculations are realistic
even down to threshold if a treatment of kinematic mass effects 
is supplemented.

Recently, results on cumulant moments have been presented by the ZEUS group at
HERA (see talk by N. Brook \cite{brook}). The moments have been
determined directly from the momentum distribution of particles in the Breit
frame. The $\xi_p$ distributions are seen to extend beyond the 
theoretical limit $Y$. The cumulant moments of order $q\geq 2$
determined from this distribution  
show large deviations from the MLLA predictions 
at low energies $Q^2$. The kinematic effects become less important at higher
energies and at $Q^2\gtrsim 1000$ GeV$^2$ the agreement with the 
predictions using $Q_0=\Lambda$ 
is restored. These results demonstrate the
importance of the soft region in the analysis of the $\xi$-moments.   
 
%

\section{Particle Spectra: the limit of small momenta $p,\ p_T\to 0$}

In this limit simple expectations follow from the coherence of the soft
gluon emission. If a soft gluon is emitted from a $q\overline q$ two jet
system then it cannot resolve with its large wave length 
all individual partons but only \lq\lq sees''
the total charge of the primary partons $q\overline q$. 
Consequently, in the analytical treatment, 
the soft gluon radiation is determined by the Born term of $O(\alpha_s)$
and one expects a nearly energy independent soft particle spectrum \cite{adkt1}.
The consequences and further predictions have been studied 
recently in more detail.

\subsection{Energy Independence} 

\begin{figure}[t]  
          \begin{center}
          \mbox{
\mbox{\epsfig{file=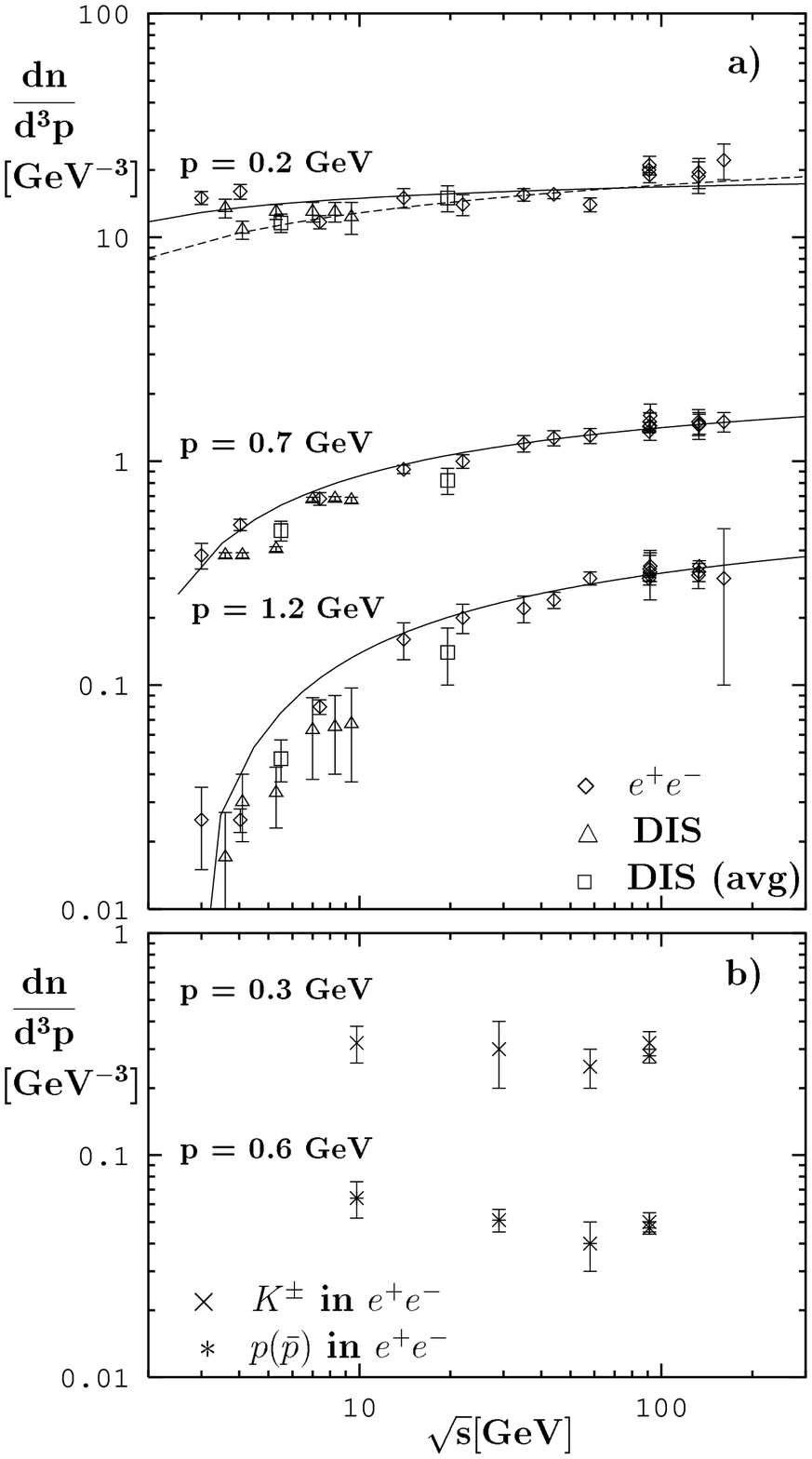,%
bbllx=2.5cm,bblly=12.6cm,bburx=16.5cm,bbury=28.0cm,%
height=6cm,clip=}} 
}          \end{center}
\vspace{-0.2cm}
\hspace{5.5cm} {\bf $\sqrt{s}\ $ GeV}
\caption{
Particle density at fixed momentum $p$ as function of $cms$ energy,
from \protect\cite{klo}
}
\label{fig:dnd3p}
\end{figure}  

The limit of small momenta $p$ and $p_T$ has been 
considered in \cite{klo}. The behaviour
of the inclusive spectrum in rapidity and for small $p_T$ is given by
\begin{equation}
\frac{dn}{dydp_T^2}\ \sim \ C_{A,F} \frac{\alpha_s(p_T)}{p_T^2} 
       \left( 1+O\left(
\ln\frac{\ln (p_T/\Lambda)}{\ln(Q_0/\Lambda)}\ 
\ln\frac{\ln(p_T/(x\Lambda))}{\ln(p_T/\Lambda)}\right)\right)
\label{Born}
\end{equation}
where the second term is known within MLLA and vanishes for $p_T\to Q_0$.
Again, the limit $p_T\to Q_0$ at the parton level 
corresponds to  $p_T\to 0$ at the hadron level. Only the first term (the Born
term) is energy independent. The approach to energy independence for the
soft particles at $p\to 0$ is seen from $e^+e^-$ data \cite{lo,klo}
and also from DIS \cite{H1}, see Fig. \ref{fig:dnd3p}. Although the detailed
behaviour depends a bit on the specific implementation of the kinematic
relations between partons and hadrons the approach towards energy independence in the limit $p\to
0$ is universal and this expectation  is nicely supported by the data.

\subsection{Colour Factors $C_A$ and $C_F$}

A crucial test of this interpretation is
the dependence of the soft particle density on the 
colour of the primary partons in  (\ref{Born}): The particle density 
in gluon and quark jets should approach the ratio
$R(g/q)=C_A/C_F=9/4$ in the soft limit.
This factor has been originally considered
for the overall event multiplicity in colour triplet and octet systems
but is approached there only at asymptotically  high energies 
\cite{bgven}. On the other hand, the prediction (\ref{Born}) 
for the soft particles applies already at finite energies \cite{klo}.

\begin{figure}[t]
          \begin{center}
  \mbox{
\mbox{\epsfig{file=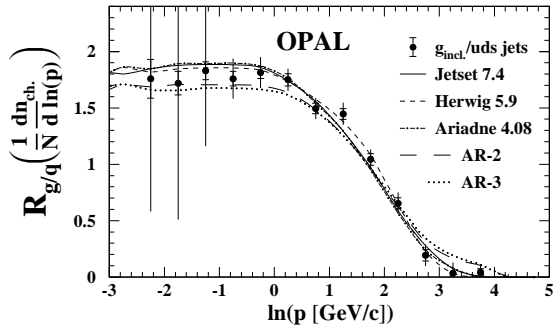,bbllx=2cm,bblly=4.2cm,bburx=19cm,bbury=14.2cm,%
height=4.5cm}}%
  }          \end{center}
\caption{
Ratio of particle densities at small momenta p in inclusive gluon jets and
quark jets \protect\cite{gary}
}
\label{fig:Rgq}
\end{figure}

In practice, it is difficult to obtain 
 $gg$ jet systems for this test. An interesting
possibility is the study of 3-jet events in $e^+e^-$ annihilation with the
gluon jet recoiling against a $q\overline q$ jet pair with relative opening
angle of $\sim 90^{\rm o}$ 
\cite{gary}. For such ``inclusive gluon jets''
 the densities of soft particles  in comparison to quark jets
approach a ratio 
$R(g/q)\sim 1.8$ for $p\lesssim 1$ GeV 
which is above the overall multiplicity ratio $\sim $ 1.5
in the quark and gluon jets but still
below the ratio $C_A/C_F=9/4$ (see Fig. \ref{fig:Rgq}). 
This difference may be attributed 
to the deviation of the events from exact collinearity.  
If the analysis is performed as function of  $p_T$ of the 
particles
the ratio 
becomes consistent with $9/4$ but not for  small $p_T<1$ GeV \cite{gary}.
This behaviour indicates the transition from the very soft emission 
which is coherent 
from all primary partons to the semisoft emission from the
parton closest in angle ($q$ or $g$) which yields directly the ratio
$C_A/C_F$.

In order to test the role of the colour of the primary partons further 
in realistic processes
it has been proposed \cite{klo} to study the soft particle emission
perpendicular to the primary partons in  3-jet events in $e^+e^-$
annihilation or in 2-jet production either in $pp$ or in $ep$ collisions,
in particular  in photoproduction. In these cases, for special limiting
configurations of the primary partons, the particle density is
either proportional to $C_F$ or to $C_A$, but it is also known for all
intermediate configurations.
 A first result of this kind of
analysis has been presented by DELPHI \cite{delphipout} which shows the
variation of the density by about 50\% in good agreement with
the prediction. The findings by OPAL \cite{gary} (Fig. \ref{fig:Rgq}) 
and DELPHI \cite{delphipout}
are  hints that also the soft particles
indeed reflect the colour charges of the primary partons.

Important tests are possible at HERA with two-jet production from 
direct and resolved photons. The former process corresponds to quark
exchange, the latter to gluon exchange. The associated soft perpendicular
radiation again reflects the different flow of the primary colour charges:
At small scattering angles $\Theta_s\to 0$ in the di-jet $cms$ 
the ratio $R_\perp$ of the soft particles 
approaches the limits
\begin{eqnarray}
{\rm direct }\; \gamma p\; {\rm production\; (q\; exchange):}&\qquad & R_\perp\to 1 \\
{\rm indirect}\; \gamma p\;
          {\rm  production\; (g\; exchange):}&\qquad & R_\perp\to C_A/C_F. 
%
\label{direct}
\end{eqnarray}

In a feasibility study \cite{bko}  using the event generator HERWIG 
these ratios have been studied as function of the particle $p_T$
and angle $\Theta_s$.
With an assumed luminosity of 4.5 $pb^{-1}$ significant results 
can be obtained. In the MC the predicted ratios are approached for small
$p_T\lesssim 0.5$ GeV but deviate considerably for larger $p_T$.
A study towards small angles $\Theta_s$ appears feasible. It would be
clearly interesting to carry out such an analysis.  

\subsection{Rapidity Plateaux}

Another consequence of the lowest order approximation (\ref{Born})
is the flat distribution in rapidity $y$ at fixed (small) 
$p_T$. An interesting
possibility appears in DIS where the soft gluon in the current hemisphere
is emitted from the quark, in the target hemisphere from a gluon. This would
lead one to expect a step in rapidity by factor $\sim$2 between both
hemispheres at high energies \cite{klo,o97}.

This problem has been studied recently by the H1 group \cite{H1plateau}.
They observed a considerable change of the rapidity spectrum with the $p_T$
cut: for large $p_T>1$ GeV the spectrum was peaked near $y=0$ in the Breit
frame -- as expected from  maximal perturbative gluon radiation  --
whereas for small  $p_T<0.3$ GeV a plateau develops in the target 
hemisphere.
On the other hand, no plateau is observed in the current direction at all.
A MC study of the $e^+e^-$ hadronic final state did not reveal a clear sign of a
flat plateau at small $p_T$ either.

The reason for the failure  seeing the flat plateau is apparently
the angular recoil of the
primary parton which is neglected in the result (\ref{Born});
this introduces an uncertainty in the definition of 
 $p_T$, especially for the higher momenta. We have investigated this 
hypothesis further by studying the rapidity distribution in 
selected MC events 
where all particles are limited in transverse momentum 
 $p_T<p_T^{max}$. Then the events are more collimated and the 
jet axis is better defined. 
The MC results in Table 1 show that the rapidity density gets flatter if
the transverse size of the jet decreases with the  $p_T^{max}$ cut
 which is in support of the above
hypothesis. This selection, however, considerably reduces the event sample.
A step in the rapidity hight of DIS events should therefore be expected only
in events with strong collimation of particles.

\begin{table}
\caption{Density of particles with  $p_T<0.15$ GeV 
in rapidity $y$, normalized at $y=-1$, 
 in events with  $p_T<p_T^{max}$ selection.
Results obtained from
the ARIADNE MC \cite{ariadne} (parameters $\Lambda=0.2$ GeV,
$\ln(Q_0/\Lambda)=0.015$ as in \cite{low}) 
}
\begin{center}
\renewcommand{\arraystretch}{1.4}
\setlength\tabcolsep{5pt}
\begin{tabular}{lllllll}
\hline\noalign{\smallskip}
$p_T^{max}$ & $y=0$ &  $y=-1$ &  $y=-2$ &  $y=-3$ & $y=-4$ & fraction of
events\\
\noalign{\smallskip}
\hline
\noalign{\smallskip}
no cut & 1.03 & 1.0 & 0.69 & 0.34 & 0.052 & 100 \% \\
0.5    & 0.9 & 1.0 & 0.78 & 0.50 & 0.13 & 9 \% \\
0.3    & 0.9 & 1.0 & 0.90 & 0.84 & 0.37 & 0.7 \% \\
\hline
\end{tabular}
\end{center}
\label{Table}
\end{table}

\subsection{Multiplicity Distributions of Soft Particles and Poissonian
Limit}

The considerations on the inclusive single particle distributions can be
generalized to multiparticle distributions
\cite{low}. Interesting predictions apply
for the multiplicity distributions of particles which are restricted in
either the transverse momentum $p_t<p_T^{cut}$ or in spherical momentum
$p<p^{cut}$.

In close similarity to QED the soft particles are independently emitted in
rapidity for limited $p_T$: because of the soft gluon coherence the
secondary emissions at small angles are suppressed. This is less so for the
spherical cut. For small values of the cut parameters one finds
the following limiting behaviour of the normalized factorial multiplicity
moments
\begin{eqnarray}
{\rm cylinder:} \qquad  & F^{(q)}(X_{\perp},Y) & \simeq \
1+\frac{q(q-1)}{6}\frac{X_{\perp}}{Y}\\
{\rm sphere:} \qquad & F^{(q)}(X,Y) & \simeq \ {\rm const}
\label{Poisson}
\end{eqnarray}
where we used the logarithmic variables $X_{\perp}=\ln(p_T^{cut}/Q_0)$, 
 $X=\ln(p^{cut}/Q_0)$ and $Y=\ln(P/Q_0)$ at jet energy $P$. Both cuts act
quite differently and for small cylindrical cut $p_T^{cut}$ the
multiplicity distribution approaches a Poisson distribution (all moments
$ F^{(q)}\to 1$). 

This prediction is verified by the ARIADNE MC at the parton
level. Interestingly, the predictions from the full  hadronic final
state after string hadronization yield factorial moments rising at small
$p_T^{cut}<1$ GeV. These predictions provide a
novel test of soft gluon coherence in multiparticle production.

\section{Conclusions and Physical Picture}
The simple idea to derive hadronic multiparticle phenomena directly
from the partonic final state works surprisingly well also for the soft
phenomena discussed here
which do not belong to the standard repertoire of perturbative QCD.
Nevertheless, some clear QCD effects can be noticed
 in the soft phenomena
and the three questions at the end of the introduction
can be answered positively. A description with small cut-off 
$Q_0\sim\Lambda$ is possible
for various inclusive quantities. The coupling is running by
more than an order of magnitude at small scales as is seen, 
in particular, in the transition from jets to hadrons.
Also, coherence effects from soft gluons are reflected in the behaviour of
soft particles. These  effects for the soft particles need further
comparison with quantitative predictions. Especially worthwhile are the tests
on soft particle flows as function of the primary emitter configuration. 
Predictions exist also for nontrivial limits of 
multiparticle soft correlations. 
 
The different threshold behaviour of partons and hadrons
can be taken into account by  appropriate relations between the 
respective kinematical
variables.
Some apparent discrepencies between MLLA predictions and observations
can be related to such mass effects.

\begin{figure}[b]
          \begin{center}
          \mbox{ 
\mbox{\epsfig{file=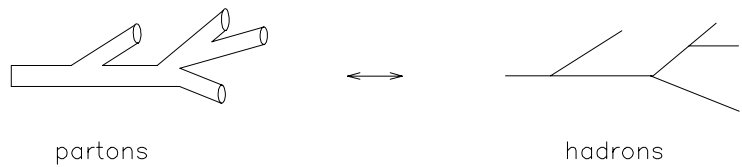,bbllx=1.5cm,bblly=16.5cm,bburx=20.0cm,bbury=21.cm,%
width=10cm}}
}          \end{center}
\vspace{-0.3cm}
\caption{
Dual picture of parton  and hadron cascades. Ultrasoft partons are
confined to 
narrow tubes with $p_T<Q_0\sim\Lambda$ around the  partons in the
perturbative cascade.
}
\label{fig:tubes}
\end{figure}

Finally, we remark on the physical picture which is supported by these
results (Fig. \ref{fig:tubes}). 
The partons in the perturbative cascade are accompagnied by ultrasoft
partons with $p_T\lesssim Q_0\sim\Lambda$ 
as in very narrow jets; they cannot be
further resolved because of confinement and therefore the perturbative
partons resolved with  $p_T\geq Q_0$
correspond to  single
final hadrons. This is consistent with the finding of normalization unity
($K=1$ in (\ref{lphdeq})) 
 in the transition jet $\to$ hadron $(y_{cut}\to 0$).
Colour at each perturbative vertex can be neutralized
by the (non-perturbative) emission of one (or several) soft quark pairs;
 in this way
the partons in the perturbative cascade evolve as  colour neutral systems
outside a volume with confinement radius $R\sim Q_0^{-1}$. 
In the timelike cascade
there is only parton splitting,
no parton recombination into massive colour singlets as in the 
preconfinement model. Such a picture can only serve as a rough guide, it can
certainly not be complete as is exemplified by the existence of resonances. 
Nevertheless, its intrinsic simplicity with only one non-perturbative
parameter $Q_0$ besides the QCD scale $\Lambda$ makes it attractive as a
guide into a further more detailed analysis.

%
\clearpage
\addcontentsline{toc}{section}{Index}
\flushbottom
\printindex


\begin{thebibliography}{8.}
\addcontentsline{toc}{section}{References}

\bibitem{dkmt2}  
 Yu. L. Dokshitzer, V. A. Khoze, A. H.
Mueller and S.\ I.\ Troyan, \emph{Basics of Perturbative
QCD}, ed. by J. Tran Thanh Van (Editions Fronti\'{e}res,
Gif-sur-Yvette, 1991)

\bibitem{ko}
 V.A.~Khoze, W.~Ochs, Int.~J.~Mod.~Phys.~A \textbf{12}, 2949 (1997) 

\bibitem{preconf}   
 D. Amati, G. Veneziano, Phys. Lett. B \textbf{83}, 87 (1979)

\bibitem{adkt1}   
 Ya. I. Azimov, Yu. L. Dokshitzer, V. A.
Khoze and S. I. Troyan, Z. Phys. C \textbf{27}, 65 (1985);
C \textbf{31}, 213 (1986)

                   

\bibitem{lo}  
S. Lupia, W. Ochs, Phys. Lett. B \textbf{365}, 339 (1996);
Eur. Phys. J. C \textbf{2}, 307 (1998)

\bibitem{klo}
V. A. Khoze, S. Lupia, W. Ochs,
Phys. Lett. B \textbf{386}, 451 (1996); 
Eur. Phys. J. C \textbf{5}, 77 (1998)
 
\bibitem{lo2}   
S. Lupia, W. Ochs, Phys. Lett. B \textbf{418}, 214 (1998)


\bibitem{durham}   
Yu. L. Dokshitzer, Proc. Durham Workshop, see
W. J. Stirling, J.\ Phys. G \textbf{17}, 1567 (1991)

\bibitem{kl}
G. Kramer, B. Lampe,            
Z. Phys. C \textbf{34}, 497 (1987); C \textbf{39} (1988)            


\bibitem{L3jmul}
 L3 Collaboration: O. Adriani et al., Phys. Lett. B \textbf{284}, 471 (1992)

\bibitem{opaljmul}
OPAL Collaboration: R.\ Acton et al., Z. Phys. C \textbf{59}, 1 (1993)

                                                                         
\bibitem{cdotw}   
 S. Catani, Yu. L. Dokshitzer, M. Olsson,                            
G.\ Turnock  B. R. Webber, Phys. Lett. B \textbf{269}, 432 (1991)           

\bibitem{Pfeifenschneider} 
P. Pfeifenschneider, thesis, Technical University Aachen, 1999;
OPAL preliminary, Physics note PN403, July 1999. 

\bibitem{FW}
C. P. Fong,  B. R. Webber, Nucl. Phys. B \textbf{355}, 54 (1991)

\bibitem{DKTInt}
Yu. L. Dokshitzer, V. A. Khoze, S. I. Troyan, Int. J. Mod. Phys. A
\textbf{7}, 1875 (1992) 

\bibitem{lupia}
S. Lupia, Phys. Lett. B \textbf{439}, 150 (1998)

\bibitem{brook}
ZEUS Collaboration: J. Breitweg et al., hep-ex/9903056 


\bibitem{H1}
 H1 Collaboration:  C. Adloff et al., Nucl. Phys. B \textbf{504}, 3  (1997)

\bibitem{bgven}
 S. J. Brodsky,  J. F. Gunion, Phys. Rev.
Lett. \textbf{37}, 402 (1976); \\
K. Konishi, A. Ukawa and G. Veneziano, 
Phys. Lett. B \textbf{78}, 243 (1978)

\bibitem{gary}
OPAL Collaboration: Abbiendi et al., hep-ex/9903027, subm. Eur. Phys. J. C.;
J. W. Gary, Phys. Rev. D \textbf{49}, 4503  (1994)

\bibitem{delphipout}
K. Hamacher, O. Klapp, P. Langefeld, M. Siebel,
DELPHI 99-115 CONF 302, subm. to the HEP'99 Conference, Tampere,
Finland, July 1999

\bibitem{bko}
J. M. Butterworth, V. A. Khoze and W. Ochs, J. Phys. 
G \textbf{25}, 1457 (1999)

\bibitem{o97}
W. Ochs, in \emph{
Proc. Ringberg Workshop `New Trends in HERA Physics',
Tegernsee, Germany 1997}, ed. by B. A. Kniehl, G. Kramer, A. Wagner (World
Scientific, Singapore, 1998) p. 173
 
\bibitem{H1plateau}
H1 Collaboration: C. Adloff et al. in:
\emph{29th Int. Conf. on High Energy Physics,Vancouver,
Canada, July 1998}, paper 531; K. T. Donovan, D. Kant, G. Thompson, J. Phys.
G \textbf{25}, 1448 (1999)

\bibitem{low}
S. Lupia, W. Ochs, J. Wosiek,  Nucl. Phys. B \textbf{540}, 405
(1999) 

\bibitem{ariadne}
L. L\"onnblad, Comp. Phys. Comm. \textbf{71}, 15 (1998)


\end{thebibliography}
\end{document}